\documentclass[twocolumn,aps,prr,showpacs,raggedbottom,nobalancelastpage,amssymb,superscriptaddress]{revtex4-2}
\usepackage[utf8]{inputenc}
\pdfoutput=1
\usepackage{color}
\usepackage{graphicx}
\usepackage{physics}
\usepackage{braket}
\usepackage{mathtools}
\usepackage[utf8]{inputenc}
\usepackage{siunitx}
\usepackage{lipsum}

\newcommand{\cdag}{c^{\dag}}

\renewcommand{\ddag}{d^{\dag}}

\newcommand{\cpdag}{c^{\phantom\dag}}

\newcommand{\veck}{\boldsymbol{k}}

\DeclareMathOperator{\sgn}{sgn}

\usepackage{tikz}

\begin{document}
\title{Unraveling screening mechanisms in Kondo impurities using an NRG-MPS-based method}

\author{Lidia~Stocker}
\affiliation{Institute for Theoretical Physics, ETH Zurich, 
	8093 Zurich, Switzerland}\thanks{Current address: Max Planck Institute for the Physics of Complex Systems, 01187 Dresden, Germany}

 \author{Oded~Zilberberg}
\affiliation{Department of Physics, University of Konstanz, 78457 Konstanz, Germany}

\begin{abstract}
The Kondo effect is a hallmark of strongly-correlated systems, where an impurity’s local degrees of freedom are screened by conduction electrons, forming a many-body singlet. With increasing degrees of freedom in the impurity, theoretical studies face significant challenges in accurately identifying and characterizing the underlying mechanisms that screen the impurity. In this work, we introduce a straightforward yet powerful methodology for identifying the formation of Kondo singlets and their screening mechanisms, by utilizing the numerical renormalization group (NRG) combined with the matrix product states (MPS) technique. We demonstrate the effectiveness of our method on the single and two-level Anderson impurity models (AIM). Furthermore, we outline generalizations to multichannel and multiorbital Kondo impurities, showing how advanced tensor network techniques render our approach a versatile framework for tackling complex impurity systems.
\end{abstract}
\maketitle
\section{Introduction}

Strongly-correlated effects in condensed matter physics unveil the profound interplay between many-body interactions and decoupling mechanisms~\cite{bruus_many-body_2004}. A paradigmatic example is the Kondo effect, where a localized impurity spin hybridizes with the surrounding conduction electrons~\cite{kondo_resistance_1964}. At high temperatures, the impurity spin is weakly coupled to the electron spins and remains unscreened. As the temperature decreases, the impurity-environment interaction becomes dominant, causing the conduction electrons surrounding the impurity to organize into a cloud that collectively screens its spin~\cite{wilson_renormalization_1975}. This cloud, known as the Kondo screening cloud, leads to the formation of a (Kondo) singlet ground state between the impurity spin and the surrounding electron spins, as illustrated in Fig.~\ref{fig: figure 1}(a).

Kondo effects are typically detected through transport measurements, such as the temperature dependence of resistivity in impurity systems~\cite{ng_-site_1988,glazman_resonant_1988,goldhaber-gordon_kondo_1998,cronenwett_tunable_1998,rossler_transport_2015,ferguson_long-range_2017,nicoli_cavity-mediated_2018,kurzmann_kondo_2021}. At high temperatures, the system exhibits conventional behavior as the impurity remains unscreened. As the temperature decreases, the formation of the Kondo cloud leads to a rise in resistivity. At low temperatures, the resistivity follows a characteristic logarithmic dependence, which is a hallmark of the Kondo effect~\cite{kondo_resistance_1964}. Beyond transport measurements, theoretical approaches have introduced the use of alternative order parameters, such as spin correlation functions~\cite{lechtenberg_equilibrium_2018}, charge fluctuations~\cite{komijani_emergent_2019}, and entanglement negativity~\cite{kim_universal_2021} to identify the impurity screening mechanisms.

Complex impurity systems, particularly those with multiple orbitals, energy levels, or environments, are known to give rise to a variety of Kondo-like states, presenting challenges for both experimental detection and theoretical modeling~\cite{affleck_critical_1991,parcollet_overscreened_1998,von_delft_2-channel_1999,paaske_non-equilibrium_2006,mora_theory_2009,mitchell_two-channel_2012,minamitani_symmetry-driven_2012,kimura_ferminon-fermi_2017,moro_quantifying_2018,blesio_fully_2019,mitchell_so5_2021,shim_hierarchical_2023,trishin_tuning_2023,chen_programmable_2024}. These systems naturally raise questions about how to identify Kondo impurities and characterize their underlying screening mechanisms. On the theoretical side, the NRG method has been the gold standard for addressing the Kondo problem~\cite{wilson_renormalization_1975}, offering precise calculations of transport, thermodynamic, spectral, and dynamic properties~\cite{oliveira_generalized_1994}. However, for complex impurity systems, conventional NRG presents significant challenges~\cite{toth_density_2008,kugler_strongly_2020,stadler_hundness_2019}. The NRG-MPS framework~\cite{saberi_matrix-product-state_2008,weichselbaum_variational_2009}, a hybrid approach combining NRG and MPS, has emerged as a promising alternative for complex impurity systems. A key limitation of NRG-MPS is its inherently ground-state-focused nature, which prevents direct access to finite-temperature properties and transport phenomena.

Recently, we proposed a NRG-MPS-based impurity tomography procedure to detect impurity-environment hybridizations as signatures of phenomena like Kondo singlet formations~\cite{stocker_entanglement-based_2022,stocker_coherent_2024}. Although our approach provides valuable insights into impurity-environment hybridization, it does not determine whether this hybridization corresponds to effective Kondo screening. Moreover, in systems with multiple environmental degrees of freedom, this method cannot identify which environmental components hybridize with the impurity.

In this work, we address these challenges by presenting a method designed to identify magnetic (Kondo) impurities and elucidate their screening mechanisms by the environment. Our approach is based on the NRG-MPS method and utilizes the fact that the MPS state offer access to nonlocal correlators. Thus, by projecting the ground state onto different subregions of the Hilbert space, we can use entanglement measures to identify the modes that coherently screen the impurity, as well as quantify the screening length. In the following, we outline the physical motivation, present the method in detail, and demonstrate its effectiveness in capturing the Kondo effect’s screening behavior in the examples of the single and two-level AIMs.

\section{Background}\label{sec: Background}
In this Section, we introduce the AIMs and their connection to Kondo models, as well as the numerical NRG-MPS algorithm underlying our method. We then describe the numerical NRG-MPS approach and review existing order parameters for detecting Kondo phenomena, emphasizing their limitations.
\subsection{Anderson impurity models}\label{sec: AIM}
\begin{figure}
\centering\includegraphics[width=8.6cm]{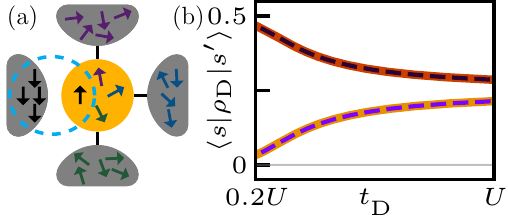}
	\caption{\textit{Anderson impurities and Kondo effects.} (a) Sketch of  a multilevel Anderson impurity Hamiltonian, cf. Eq.~\eqref{eq: hamiltonian anderson impurity multilevel multiorbit}. The impurity (yellow circle) contains multiple levels that are tunnel-coupled to as many multiple environments. In the single AIM with a single level and environment, a Kondo singlet can form between impurity and environment (dashed blue circle). (b) NRG-MPS tomography results of the single AIM, where we highlight the impurity's reduced density matrix elements $\bra{\uparrow} \rho_\text{imp} \ket{\uparrow}$ (black), $\bra{\downarrow} \rho_\text{imp} \ket{\downarrow}$ (red), $\bra{\uparrow\downarrow} \rho_\text{imp} \ket{\uparrow\downarrow}$ (orange), and $\bra{0} \rho_\text{imp} \ket{0}$ (purple) cf. Eq.~\eqref{eq: reduced density matrix Kondo single imp} and Refs.~\cite{stocker_entanglement-based_2022,stocker_coherent_2024}. The other matrix elements $\bra{s} \rho_\text{imp} \ket{s'}$ are negligible (thin grey lines). We consider the impurity at half filling $\epsilon_\text{D} = -U/2$, set the Wilson chain length $N_\text{chain} = 90$ with MPS bond dimension $D = 100$, and assume the environment to have a constant density of states $d_0 = 1/(2U)$.}
	\label{fig: figure 1}
\end{figure}

The single AIM is one of the most extensively studied quantum impurity systems~\cite{anderson_localized_1961}. It describes an interacting localized impurity that is tunnel-coupled to an environment, i.e., a surrounding conduction band. Generalizations of the single AIM include coupling of the impurity to multiple environments (or environmental degrees of freedom) and impurities with multiple orbitals and levels. These generalizations effectively describe the intricate interactions manifesting in realistic materials with spin, multiband structures, and orbital degrees of freedom. In this work, we consider a multilevel AIM of the form
\begin{align}\label{eq: hamiltonian anderson impurity multilevel multiorbit}
  	{H}  = & \underbrace{\sum_\ell^{\phantom \dag}\epsilon_{\ell}^{\phantom\dag}n_{\ell\sigma}^{\phantom\dag}+\sum_{\ell\sigma\neq\ell'\sigma'}U_{\ell\ell'}^{\phantom\dag}n_{\ell\sigma}^{\phantom\dag}n_{\ell'\sigma'}^{\phantom\dag}}_{H_\text{imp}} + \underbrace{\sum_{\ell\sigma k}^{\phantom\dag}\epsilon_{\ell \mathbf{k}}^{\phantom\dag}\cdag_{\ell \mathbf{k}\sigma}\cpdag_{\ell \mathbf{k}\sigma}}_{H_\text{env}} \nonumber \\
  	& + \underbrace{\sum_{\ell \mathbf{k}\sigma}t_{\ell \mathbf{k}}^{\phantom\dag}\ddag_{\ell\sigma} \cpdag_{\ell \mathbf{k}\sigma}+\text{H.c.}}_{H_\text{tun}^{\text{imp-env}}} \ ,
\end{align}
see Fig.~\ref{fig: figure 1}(a). Here, $n_{\ell\sigma}^{\phantom\dag} = d_{\ell\sigma}^{\dag}d_{\ell\sigma}^{\phantom\dag}$ is the occupation operator of the $\ell=1,\ldots,L$ impurity level with $d^\dag_{\ell\sigma}, d^{\phantom\dag}_{\ell\sigma}$ the fermionic creation and annihilation operators with spin $\sigma \in \{\uparrow, \downarrow\}$. The operators $c^\dag_{\ell\veck\sigma}, c^{\phantom\dag}_{\ell\veck\sigma}$ act similarly for the conduction electrons in the environment with momentum $\mathbf{k}$ and spin $\sigma$. Note that we assume the impurity and environment to have the same number of levels. Each impurity level has energy $\epsilon_{\ell}$, the levels interact with $U_{\ell\ell'}$ as long as Pauli's exclusion principle holds, the environment energies $\epsilon_{\ell \mathbf{k}}$ are counted on each level, and the tunnel coupling is level selective $t_{\ell \mathbf{k}}$. Throughout this work, we mostly employ the single-level AIM, where $\ell=\text{D}$ and we label $\epsilon_\text{D}, U, \epsilon_{\veck}, t_{\text{D}\veck}$ the onsite energy, Coulomb repulsion, environment's energy levels and impurity-environment couplings.

\subsection{Kondo models}\label{sec: Kondo models}
The Kondo model involves an alternative effective Hamiltonian,
 \begin{equation}
     H =  H_\text{env}^{\phantom {i}} + H_\text{int}^{\text{imp}-\text{env}} \ ,
 \end{equation}
 to describe the strong hybridization of the impurity with the environment, by focusing on spin scattering from a magnetic impurity~\cite{kondo_resistance_1964}. This scattering dominates in situations where the impurity is occupied by a single electron. Generalizations of the Kondo model include systems with multiple environments and multiorbit impurities~\cite{nozieres_kondo_1980,cox_exotic_1998,parcollet_overscreened_1998}. Similarly to the discussion of the Anderson impurity model, here we consider a generalization to multilevel impurities, where the impurity-environment interaction term reads
\begin{multline}
   H_\text{int}^{\text{imp}-\text{env}}  = \sum_\ell H_\text{int}^{\text{imp}_\ell-\text{env}} =  \sum_{\ell'\ell''}\frac{J_{\ell'\ell''}^\ell}{2}\cdot\\\underbrace{\left(\sum_{\sigma\sigma'}d_{\ell\sigma}^\dag \boldsymbol{\tau}_{\sigma\sigma'}^{\phantom\dag}d_{\ell\sigma'}^{\phantom\dag}\right)}_{2\boldsymbol{S}_{\text{imp}_{\ell}}}\cdot\left(\sum_{\veck\veck'}\boldsymbol{S}_{\ell'\veck,\ell''\veck'}\right) \ , \label{eq: hamiltonian kondo model multilevel multiorbit} 
\end{multline}
where $J_{\ell'\ell''}^{\ell}$ is the magnetic interaction term between the $\ell^{\rm{th}}$ impurity level and the environment, $\boldsymbol{S}_{\text{imp}_\ell}$ and $\boldsymbol{S}_{\ell\veck,\ell'\veck'}$ are the spin operators of the $\ell^{\text{th}}$ impurity level and environment with $\ell,\ell'$ degrees of freedom. Additionally, $\boldsymbol{\tau}_{\sigma,\sigma'}$ is the Pauli matrix vector, assuming $\hbar=1$. The singly-occupied impurity levels have two-degenerate states that do not contribute to the effective energy, thus, $H_\text{imp}$ is removed. The single-level Kondo model, Eq.~\eqref{eq: hamiltonian kondo model multilevel multiorbit} with $\ell=\text{D}$, can be derived from the single AIM in the singly-occupied regime, Eq.~\eqref{eq: hamiltonian anderson impurity multilevel multiorbit} with $\ell=\text{D}$, using a Schrieffer-Wolff transformation~\cite{schrieffer_relation_1966, bruus_many-body_2004}. This transformation eliminates charge fluctuations and retains only the spin degrees of freedom. Additionally, the transformation determines the Kondo coupling in terms of the system's coupling parameters
\begin{equation}
    J = \frac{2U\left|t_\text{D}\right|^2}{\left(\epsilon_\text{D} + U\right)\left(-\epsilon_\text{D}\right)} \ .
\end{equation}
At half-filling $\epsilon_\text{D} \approx - U/2$, the coupling becomes antiferromagnetic with $J > 0$. Therefore, an impurity-environment Kondo singlet can emerge in the single AIM at low temperatures~\cite{jones_low-temperature_1988, glazman_resonant_1988, ng_-site_1988, kawabata_electron_1991}. The critical (Kondo) temperature of this system is estimated as~\cite{schrieffer_relation_1966,bruus_many-body_2004}
\begin{equation}\label{eq: Kondo temperature}
    k_\text{B}T_\text{K}\approx \sqrt{\frac{\pi\left|t_\text{D}\right|^2Ud_0}{2}}\exp\left(\frac{\epsilon_\text{D}\left(\epsilon_\text{D}+U\right)}{2\left|t_\text{D}\right|^2Ud_0}\right) \ .
\end{equation}
The Kondo singlet state can be expressed as
\begin{equation}\label{eq: Kondo SU(2) singlet}
    \ket{\psi} = \frac{1}{\sqrt{2}}\left(\ket{\uparrow}_\text{imp}\otimes\ket{\downarrow}_\text{env} - \ket{\downarrow}_\text{imp}\otimes\ket{\uparrow}_\text{env}\right) \ ,
\end{equation}
where $\ket{\sigma}_\text{imp}$ represents the state of the impurity spin and $\ket{\sigma}_\text{env}$ represents the projected spin of the environment state. Due to the two-fold degeneracy of the impurity level, the system exhibits $SU(2)$ symmetry, and the strong impurity-environment hybridization is dubbed $SU(2)$ Kondo.

In multi-level systems, $\ell > 1$, $N \geq 2$ degenerate spin and orbital states are present and the electron spins of the different impurity levels are screened by the electrons of the environment, similarly to the standard $SU(2)$ Kondo model. At temperatures below the critical Kondo temperature, the impurity and environment form a many-body ground state
\begin{equation}\label{eq: Kondo state impurity plus environment multilevels}
    \ket{\psi} = \frac{1}{\sqrt{\sum_{\ell\ell'\sigma}\left|a_{\ell\ell'\sigma}^{\phantom *}\right|^2}}\sum_{\sigma\ell\ell'}a_{\ell\ell'\sigma}^{\phantom *}\ket{\sigma}_{\text{imp}_\ell}\otimes\ket{\bar{\sigma}}_{\text{env}_{\ell'}} \,,
\end{equation}
where $a_{\ell\ell'\sigma}^{\phantom *}$ are general prefactors, of which $N$ are nonzero. Notably, the amplitudes $\left|a_{\ell\ell'\sigma}\right|$ of these coefficients are $\sigma$-independent, corresponding to a spin degeneracy in the impurity's levels, which is required for the Kondo effect to emerge. In configurations where $N$ degenerate impurity states are screened by the environment and the Kondo coupling is $\ell$-independent $J_{\ell\ell'}\equiv J$, this phenomenon is referred to as the $SU(N)$ Kondo effect. This corresponds also to $\ell$ amplitude-independent coupling coefficients $\left|a_\ell\right|\equiv \left|a\right|$. Note that a Schrieffer-Wolff transformation can also be applied to a multilevel AIM, cf. Eq.~\eqref{eq: hamiltonian anderson impurity multilevel multiorbit}. In this case, the system maps to a multilevel Kondo model, where $J_{\ell'\ell''}^\ell = 0$ for $\ell \neq \ell' \neq \ell''$. Consequently, the $\ell^\text{th}$ impurity level can only be screened by the $\ell^\text{th}$ environment level.

\subsection{NRG-MPS method}\label{sec: NRG-MPS method}
 In this Section, we provide a summary of the necessary ingredients of the NRG-MPS method that form the foundation for our proposed procedure in Sec.~\ref{sec: method}. As mentioned earlier, the Kondo effect has been both theoretically and experimentally identified through transport measurements. One prominent non-perturbative method for calculating transport through quantum impurity systems, which captures strongly-correlated effects, is the NRG method~\cite{wilson_renormalization_1975, krishna-murthy_renormalization-group_1980, krishna-murthy_renormalization-group_1980-1}. In the NRG approach, the environment, e.g. $H_\text{env}$ in Eq.~\eqref{eq: hamiltonian anderson impurity multilevel multiorbit}, is mapped onto a \textit{Wilson chain}, i.e., the conduction band states are discretized and sampled, such that they can be effectively written as a one-dimensional semi-infinite chain. The ground state of the Wilson chain provides an accurate description of low-energy states of the impurity-environment system. For the  single AIM, the Wilson chain of the full system is given by
\begin{align} \label{eq: hamiltonian AIM wilson chain}
	H = & H_{\text{imp}} + \sum_{\sigma, n=0}^{\infty} \epsilon_{n} c_{n\sigma}^\dagger c_{n\sigma} + \sum_{\sigma} t_\text{D}^{\phantom\dag}\left(d_{\sigma}^\dagger c_{0\sigma}^{\phantom\dag} + \text{H.c.}\right)  \nonumber \\
	& + \sum_{\sigma, n=0}^{\infty} t_{n}^{\phantom\dag} \left(c_{n\sigma}^\dagger c_{n+1\sigma}^{\phantom\dag} + \text{H.c.}\right),
\end{align}
where $c_n^\dagger$ and $c_n$ are the creation and annihilation operators at the $n^\text{th}$ site of the Wilson chain~\cite{bulla_numerical_2008}. The energy and hopping terms, $\epsilon_n$ and $t_n$, can be determined recursively, see Appendix~\ref{App: Wilsons transformation results}. Assuming energy-independent coupling amplitudes $t_{\veck} = t$ and a constant density of states for the environment, the hopping terms decay as
\begin{equation}
\label{eq: hopping terms wilson chain}
	t_{n} \sim \Lambda^{-n/2}\ ,
\end{equation}
where $\Lambda$ is the discretization parameter that controls the energy resolution of the sampling~\cite{krishna-murthy_renormalization-group_1980,krishna-murthy_renormalization-group_1980-1}. 

In the general case of multilevel impurity systems, cf. Eq.~\eqref{eq: hamiltonian anderson impurity multilevel multiorbit}, each environment level is similarly transformed into a Wilson chain, resulting in an impurity coupled to $L$ distinct one-dimensional chains~\cite{mitchell_generalized_2014}. The NRG method for quantum impurities systematically further reduces the dimension of the system's Hilbert space through an iterative diagonalization procedure, progressively adding subsequent chain sites $n=0,1,\ldots$. Initially introduced as the first solution to the Kondo problem~\cite{wilson_renormalization_1975}, NRG has been widely successful in studying a variety of quantum impurity systems~\cite{pruschke_low-energy_2000,mitchell_generalized_2014}. However, it encounters significant limitations when applied to multiorbital or multilevel systems, as well as to impurities coupled to multiple environments~\cite{toth_density_2008,kugler_strongly_2020,stadler_hundness_2019}.

In this context, the NRG-MPS approach offers a promising alternative~\cite{saberi_matrix-product-state_2008, weichselbaum_variational_2009}. In the NRG-MPS method, the environment is also transformed into a Wilson chain, cf.~Eq.~\eqref{eq: hamiltonian AIM wilson chain}. This transformed environment is truncated at a finite length $N_\text{chain}$. This truncation is justified by the exponential decay of the hopping terms, cf.~Eq.~\eqref{eq: hopping terms wilson chain}. As a result, the impurity system is approximated by a finite one-dimensional chain, making it highly compatible with tensor network MPS studies. 

The MPS formalism enables efficient representation of quantum states and operators by truncation of irrelevant degrees of freedom in the Hilbert space~\cite{schollwock_density-matrix_2011}. These truncations reduce the system's dimensionality from $\mathcal{O}(d^{N_\text{chain}})$ to $\mathcal{O}(d{N_\text{chain}} D)$, where $d$ is the dimension of the local Hilbert space of each chain site and $D$ is the so-called \textit{bond dimension}. The latter is the truncation parameter of the MPS method that is typically taken on the order of $O(10^2)$. Within the MPS framework, the Hamiltonian is represented as a matrix product operator (MPO), and the many-body ground state of the system, including both the impurity and its environment, is calculated using the density matrix renormalization group (DMRG) algorithm~\cite{white_density_1992,white_density-matrix_1993}. This approach provides a non-perturbative solution and has been shown to overcome some of the limitations of the NRG method, particularly for multienvironment or multilevel systems~\cite{saberi_matrix-product-state_2008, weichselbaum_variational_2009}.

\subsection{Tomography on NRG-MPS impurities}\label{sec: Tomography on NRG-MPS impurities}
While the NRG-MPS method excels in ground-state calculations, it falls short in addressing transport observables, such as conductance or resistivity, which are key indicators of the Kondo effect. Hence, extensions of this ground-state-focused approach are needed in order to describe strongly-correlated states like the Kondo effect, i.e., it requires new order parameters that can effectively describe such phenomena.
In previous works~\cite{stocker_entanglement-based_2022, stocker_coherent_2024}, we introduced the impurity's reduced density matrix as a collective order parameter for identifying the formation of strongly correlated effects in quantum impurity systems. Specifically, in the case of an impurity-environment singlet formation, as described by Eq.~\eqref{eq: Kondo SU(2) singlet}, the reduced density matrix of the impurity is given by
\begin{equation}\label{eq: reduced density matrix Kondo single imp} 
    \rho_\text{imp} = \frac{1}{2}\sum_\sigma \ket{\sigma}\bra{\sigma} \ . 
\end{equation}
In Figure~\ref{fig: figure 1}(b), we illustrate an application of the tomography procedure in the example of the single AIM. We calculate the impurity's reduced density matrix by first finding the many-body ground state using the NRG-MPS method, see Sec.~\ref{sec: NRG-MPS method}, and then tracing out the environment. At low impurity-environment coupling, the system corresponds to the configuration described by Eq.~\eqref{eq: reduced density matrix Kondo single imp}. Thus, we capture an impurity-environment singlet formation. However, as the coupling strength increases to $t_\text{D} \sim U$, this configuration becomes partially suppressed when the many-body singlet state starts to coexist with the empty and doubly-occupied impurity states. For couplings below $t_\text{D} < 0.2U$, our numerical truncations hinder the convergence into the Kondo-singlet ground state.

While the impurity tomography procedure is useful for detecting strong correlations between the impurity and its environment, it cannot distinguish between a simple two-particle singlet and more complex many-body singlet states, such as Kondo states. Additionally, in systems with multiple environments (or multiorbital/multilevel environments), it does not identify which environment degree of freedom is involved in the hybridization. These limitations highlight the need for alternative order parameters to describe the formation of Kondo impurities. It is important to note that two-point local correlators are inadequate for describing the Kondo effect, as they fail to capture the extended screening cloud formed by the conduction electrons. Hence, to accurately describe the nonlocal correlations forming in the Kondo singlet state, global correlators are required. Such correlators should account for interactions between the impurity and the entire environment.

\section{Method}\label{sec: method}
In this Section, we present a method to determine whether many-body screening occurs in a quantum impurity based on a NRG-MPS ground state calculation. Crucially, the NRG-MPS method yields an approximate description of the full ground states while keeping a separation between system and environment degrees of freedom. On top of such a representation, our method is designed to identify which of the environment degrees of freedom are coherently involved in the screening process. We present an example focused on capturing $SU(N)$ Kondo effects in multilevel systems. Crucially, our approach can be generalized to more complex systems with, e.g., multiple channels and/or orbitals, cf.~discussion in Sec.~\ref{sec: conclusion and outlook}.

As described in Sec.~\ref{sec: NRG-MPS method}, we translate the considered quantum impurity model to a Wilson chain and find the ground state using NRG-MPS. On the resulting ground state of the full system, we apply a projector operator onto the sub-Hilbert space where the $\ell^\text{th}$ level of the impurity is filled with one electron with spin $\sigma$. This is accomplished using the projection operators
\begin{equation}\label{eq: projector operators spin subhilbert spaces}
	{P}^{\sigma}_{\ell} \equiv \ket{\sigma}\bra{\sigma}_{\ell}\otimes \prod_{\ell'\neq\ell}I_{\ell'}\otimes{I}_\text{env} \ ,
\end{equation}
where ${I}_\text{env}$ ($I_{\ell'}$) is the identity operator acting on the environment ($\ell'$) subspace. With these operators, we evaluate the correlator of the $\ell^\text{th}$ level
\begin{equation}\label{eq: total projected spin impurity}
\mathcal{M}_{\ell} = \sum_\sigma\frac{1}{\bra{\sigma}{S}_{\ell}^z\ket{\sigma}_\ell}\bra{\psi} {P}^{\sigma}_{\ell}{S}_{\ell}^z{P}^{\sigma}_{\ell}\ket{\psi} \ ,
\end{equation}
where ${S}_\ell^z$ is the spin operator of the $\ell^\text{th}$ level of the impurity. Note that, by summing over all levels, the impurity's correlator $\mathcal{M}_\text{imp}=\sum_\ell\mathcal{M}_{\ell}$ returns $1$ if the impurity is in a superposition/mixture of the $\ket{\uparrow}_\text{imp}$ and $\ket{\downarrow}_\text{imp}$ states. To determine whether the environment, or the environment's levels, are screening these system's magnetic states, we similarly calculate the environment's correlator of the $\ell^{'\text{th}}$ level
\begin{equation}\label{eq: total projected spin environment}
	\mathcal{M}_{\text{env}_{\ell'}}^{\ell} = \sum_{\sigma}\frac{1}{\bra{\sigma}{S}_\ell^z\ket{\sigma}_\ell}\bra{\psi} {P}^{\sigma}_\ell{S}_{\text{env}_{\ell'}}^z{P}^{\sigma}_\ell\ket{\psi}\ ,
\end{equation}
which is the projected spin of the $\ell^{'\text{th}}$ environment level in the Hilbert subspaces obtained by applying the projector operator for the $\ell^\text{th}$ impurity level  as in Eq.~\eqref{eq: projector operators spin subhilbert spaces}.
The sum over the environment's levels and impurity's projection levels $ \mathcal{M}_{\text{env}}=\sum_{\ell\ell'}\mathcal{M}_{\text{env}_{\ell'}}^\ell $ returns $-1$ if the state of the total system is in a superposition of configurations where the impurity has projected spin $\sigma$ and the environment has spin $\bar{\sigma}$. For an approximate impurity in the Fock-1 block, where $ \mathcal{M}_\text{imp} < 1 $, an environment's correlator $\mathcal{M}_\text{env} \approx -\mathcal{M}_\text{imp}$ indicates that the environment is effectively screening the impurity. To determine which environment level is screening the spin on the $\ell^\text{th}$ level of the impurity, it is sufficient to compare $\mathcal{M}_{\ell}$ with the $\mathcal{M}_{\text{env}_{\ell'}}^\ell$ correlators.

As previously discussed, Kondo effects are established when impurity states are degenerate. It is important to note that our correlators $\mathcal{M}_{\text{imp}}$ and $\mathcal{M}_{\text{env}}$ are focused solely on the magnetic orientation/screening per level $\ell$ and are not sensitive to the degeneracy, i.e., whether a coherent superposition of screening occurs. For example, if the full system is in the product state
\begin{equation}\label{eq: nonkondo state}
    \ket{\psi} = \ket{\uparrow}_\text{imp}\otimes\ket{\downarrow}_\text{env} \ ,
\end{equation}
we will also have $\mathcal{M}_{\text{imp}} = -\mathcal{M}_{\text{env}} = 1$. Hence, to determine if the system is indeed in a Kondo configuration as in Eq.~\eqref{eq: Kondo state impurity plus environment multilevels}, we perform an additional step. First, we project the full system into the Fock-1 block of each of the impurity levels
\begin{align}\label{eq: fock-1 projected impurity s}
    \ket{\psi^{n_{\ell}s=1}} & = \frac{1}{\sqrt{\eta}}\sum_{\ell\sigma} P_\ell^\sigma\ket{\psi} \ ,
\end{align}
where $\eta$ is the normalization coefficient and with $n_\ell s=1$, we highlight the projection on each impurity level. With this projection, the projected impurity's Hilbert space is $2L$-dimensional. Next, we perform a Schmidt decomposition between the impurity and environment subspaces of the projected state. The general form of this decomposition reads
\begin{align}
    \ket{\psi^{n_\ell s=1}} & = \sum_{i=1}^{2L}\lambda_i\ket{\phi_i}_\text{imp}\otimes\ket{\phi_i}_\text{env} \ , \label{eq: schmidt decomposition imp-env}
\end{align}
where ${\ket{\phi_i}_\text{imp}}$ and ${\ket{\phi_i}_\text{env}}$ form orthonormal sets in the respective Hilbert spaces and $\lambda_i\geq0$ are the Schmidt coefficients. In a product state as in Eq.~\eqref{eq: nonkondo state}, only one Schmidt coefficient is nonzero. If the system is instead in a $SU(L)$ Kondo singlet configuration, cf.~Eq.~\eqref{eq: Kondo state impurity plus environment multilevels}, with $a_{\ell\ell'}=a$, we will have $\lambda_i \approx 1/\sqrt{2L}$ for $i=1,\ldots,L$. Thus, by combining our correlators with the analysis of Schmidt coefficients, we can determine if the full system is in a superposition of orthogonal states corresponding to a Kondo-like hybridization.

 The calculation of the $\mathcal{M}_\ell $ and  $ \mathcal{M}_{\text{env}_{\ell'}}^\ell$ correlators is particularly efficient in combination with the NRG-MPS method. This efficiency stems from the fact that it only requires the application of local projectors and operators, which require only multiplications of local tensors in MPS formalism. For a detailed discussion of these operations and the MPS formalism in general, we refer the reader to Ref.~\cite{schollwock_density-matrix_2011}. In the MPS chain, the total magnetic moment of the environment is the sum over all Wilson chain sites
\begin{equation}
	\mathcal{M}_{\text{env}_{\ell'}}^\ell= \sum_{n, n\notin\text{imp}} \mathcal{M}_{n,\ell'}^\ell \ ,\label{eq: spin projected space environment as sum of wilson chain sites}
\end{equation}where
\begin{equation}\label{eq: total projected spin environment site}
\mathcal{M}_{\text{n},\ell'}^\ell =  \sum_{\sigma}\frac{1}{\bra{\sigma}{S}_\ell^z\ket{\sigma}_\ell}\bra{\psi}P^{\sigma}_\ell S_{n,\ell'}^{z}{P}^{\sigma}_\ell\ket{\psi}  \ ,
\end{equation}
is the correlator of the $n^\mathrm{th}$ site of the $\ell^\text{th}$ level of the NRG-MPS chain, excluding the sites belonging to the impurity. The Schmidt coefficients of the impurity-environment bipartition are the square roots of the eigenvalues of the impurity's reduced density matrix. This reduced density matrix is efficiently calculated in an MPS chain~\cite{schollwock_density-matrix_2011,stocker_entanglement-based_2022}. Note that a similar projection can be done of orbital degenerate states (different $\ell$) to resolve orbital/charge Kondo screening channels. As such, we have all the ingredients to determine screening in quantum impurity systems.

\section{Results}\label{sec: Results}
As a demonstration of the method proposed in Sec.~\ref{sec: method}, we calculate the $\mathcal{M}_{\ell}, \mathcal{M}_{\text{env}_{\ell'}}^{\ell}$ correlators in both the single- and two-level AIM and analyze the screening mechanisms.
\begin{figure*}
\centering\includegraphics[width=17.4cm]{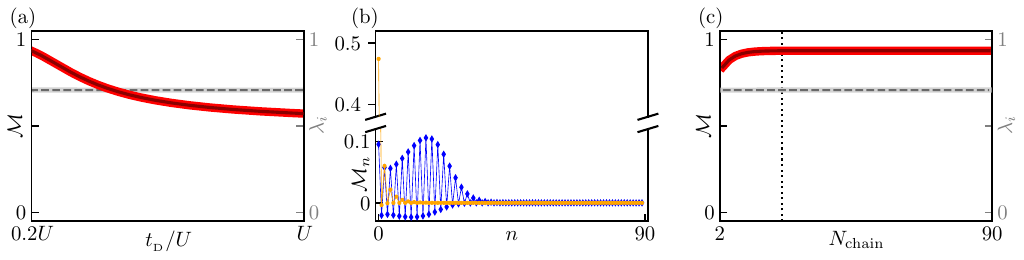}
	\caption{\textit{Screening of Kondo impurities in the single AIM~\eqref{eq: hamiltonian anderson impurity multilevel multiorbit}.} (a) Impurity $\mathcal{M}_\text{imp}$ (thick red) and environment $\mathcal{M}_\text{env}$ (thin dark red) correlators, cf.~Eqs.~\eqref{eq: total projected spin impurity} and~\eqref{eq: total projected spin environment}. Additionally, we plot the Schmidt coefficients at the impurity-environment bipartition $\lambda_1$ (light grey) and $\lambda_2$ (dashed dark grey) as defined in~Eq.~\eqref{eq: schmidt decomposition imp-env}. (b) Correlator along the Wilson chain sites $n=0,\ldots,N_\text{chain}$, cf. Eq.~\eqref{eq: total projected spin environment site} for $t_\text{D} = 0.2U$ (blue diamonds) and $t_\text{D} = U$ (orange circles). (c) Same as (a) for even Wilson chain lengths $N_\text{chain}$. Dashed vertical line indicates the value $N_\text{chain}=22$ discussed in the text. Unless otherwise specified, we use $t_\text{D} = 0.2U$, set the impurity's onsite energy $\epsilon_\text{D} = -U/2$, the Wilson chain length $N_\text{chain}=90$ and discretization parameter $\Lambda = 2$, as well as an MPS bond dimension $D=100$. We assume the environment's levels to have a constant density of states $d_0 = 1/(2U)$.}
	\label{fig: figure 2}
\end{figure*}

\subsection{Single AIM}\label{sec: Single AIM}
We apply our method to the single AIM in the  $\epsilon_\text{D} = -U/2$ regime, where the impurity is singly occupied, cf.~Eq.~\eqref{eq: hamiltonian anderson impurity multilevel multiorbit}. Our analysis focuses on quantifying the screening of the impurity by the environment. In Fig.~\ref{fig: figure 2}(a), we compare the correlator of the impurity~\eqref{eq: total projected spin impurity} with that of the full environment~\eqref{eq: total projected spin environment} as we vary the impurity-environment coupling $t_{\text{D}}$. For low coupling $t_{\text{D}} \approx 0.2U$, we observe that $\mathcal{M}_\text{imp} \to 1$ and $\mathcal{M}_\text{env} \approx -\mathcal{M}_\text{imp}$, indicating nearly perfect screening of the impurity’s magnetic moment by the environment. As the coupling increases, the impurity's magnetic moment decreases, while the environment continues to fully screen the impurity. We attribute the reduction in the impurity magnetic moment to increasing coexistence with the empty and doubly occupied impurity configurations, as revealed by the NRG-MPS tomography, see Fig.~\ref{fig: figure 1}(a). Verifying that the screening is coherent, we observe two equivalent Schmidt coefficients as in Eq.~\eqref{eq: schmidt decomposition imp-env} $\lambda_1 \approx \lambda_2 \approx 1/\sqrt{2}$  regardless of $t_{\rm D}$, see Fig.~\ref{fig: figure 2}(a). This indicates that the projected impurity state exists in a coherent superposition of the two spin configurations even when the full state of the impurity starts to mix with the empty and doubly-occupied states. Therefore, our method accurately captures the perfect environment's screening for low coupling, which hint at a Kondo singlet formation [cf.~Eq.~\eqref{eq: Kondo SU(2) singlet}].

Within the NRG-MPS framework, the correlator $\mathcal{M}_n$ of the environment~\eqref{eq: total projected spin environment site} serves as a tool for identifying the many-body nature of the hybridized impurity-environment state, see Fig.~\ref{fig: figure 2}(b). In particular, for weak impurity-environment coupling $t_{\text{D}} = 0.2$, the nonzero values of $\mathcal{M}_n$ across multiple chain sites indicate the formation of a Kondo singlet rather than a simple two-particle singlet. In contrast, when $t_{\text{D}} = U$, the first site of the Wilson chain plays the dominant role in screening, signaling that the screening does not involve long-range hybridization with the environment. Crucially, this distinct behavior cannot be resolved by the tomography procedure, cf. Fig.~\ref{fig: figure 1}(b) and Ref.~\cite{stocker_entanglement-based_2022}. Hence, we demonstrate the necessity of the nonlocal environment correlators introduced here. The values of $\mathcal{M}_n$ characteristic spatial oscillations that reflects the staggered component of the bath spin correlations. This behavior originates from the $2k_\text{F}$, where $k_\text{F}$ is the Fermi wavevector,
 contribution to the spin density, which dominates the long-distance response of a gapless environment.~\cite{friedel_xiv_1952,affleck_friedel_2008}. Additionally, in the Kondo regime, the $\mathcal{M}_n$ grow before eventually decaying to zero. This behavior originates from the Wilson transformation and the underlying properties of the Kondo state, as discussed in Appendix~\ref{App: correlator along the Wilson chain}.

Additionally, we use the quantity $\mathcal{M}_n$ to assess convergence. The NRG-MPS method relies on the fact that sites beyond a certain length in the chain (lower environment energy states) only marginally impact the Kondo screening. If $\mathcal{M}_n$ does not vanish for sites near the end of the Wilson chain, then sites with $n > N_\text{chain}$ may still influence the many-body physics of the quantum impurity and should be considered in the analysis. Otherwise, they can be truncated and Kondo physics will manifest also for shorter chains. Importantly, the Wilson chain length $N_\text{chain}$ is directly related to the system’s effective temperature
\begin{equation}
\label{eq: kondo temperature NRG}
    k_\text{B}T = \bar{\beta}^{-1}\Lambda^{-(N_\text{chain}-1)/2} \ ,
\end{equation}
where $\bar{\beta} \in\left[0.5,1\right]$~\cite{bulla_numerical_2008}. As such, we can discuss the critical temperature, i.e. the Kondo temperature $T_\text{K}$, for forming the many-body impurity-environment singlet. With the coupling parameters as in Fig.~\ref{fig: figure 2} and $t_\text{D} = 0.2U$, we obtain $k_\text{B}T_\text{K}\sim 3{\times}10^{-4}U$, cf.~Eq.~\eqref{eq: Kondo temperature}. From Eq.~\eqref{eq: kondo temperature NRG}, we obtain a minimal Wilson chain length of $N_\text{chain}^\text{min}\approx22$. Interestingly, we observe a convergence in our correlators $\mathcal{M}_\text{imp},\mathcal{M}_\text{env}$ for Wilson chain length $N_\text{chain} \geq N_\text{chain}^\text{min}$, see Fig.~\ref{fig: figure 2}(c). 
Using our correlator, we detect the formation of a Kondo impurity in the single AIM, in strong agreement with the energy scales predicted by the Schrieffer-Wolff transformation.

\begin{figure}
\centering\includegraphics[width=8.6cm]{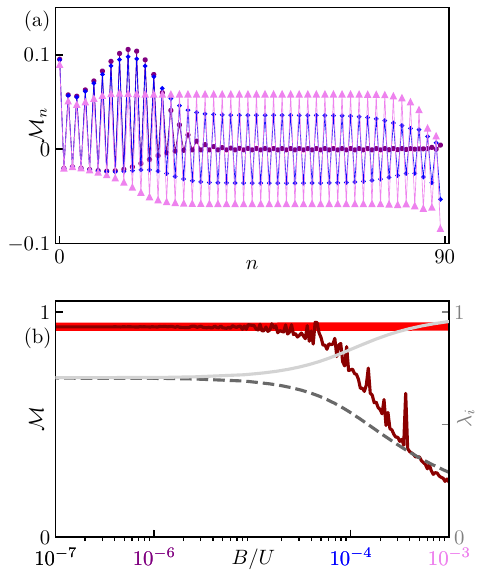}
	\caption{\textit{Screening of Kondo impurities in the single AIM~\eqref{eq: hamiltonian anderson impurity multilevel multiorbit}.} (a) Correlator of the Wilson chain sites $n=0,\ldots,N$, cf. Eq.~\eqref{eq: total projected spin environment site} for $B/U =10^{-6},10^{-4}, 10^{-3}$ (purple circles, blue diamonds, and pink triangles, respectively). (b) Impurity $\mathcal{M}_\text{imp}$ (thick red line), cf.~Eq.~\eqref{eq: total projected spin impurity}, and environment $\mathcal{M}_\text{env}$ (thin dark red line), cf. Eq.~\eqref{eq: total projected spin environment}, correlators. The Schmidt coefficients $\lambda_1$ (light grey) $\lambda_2$ (dark grey) as in Eq.~\eqref{eq: schmidt decomposition imp-env}. Colors of the x-axis labels correspond to the values as in subfigure (a). We set the impurity's onsite energy $\epsilon_\text{D} = -U/2$, $t_\text{D} = 0.2U$, and the other parameters as in Fig.~\ref{fig: figure 2}.}
	\label{fig: figure 3}
\end{figure}

\subsection{Magnetic field on a single AIM}
As a second example, we introduce a magnetic field to the single AIM. As discussed in Sec.~\ref{sec: Kondo models}, the Kondo effect arises when the impurity exhibits a degeneracy in its energy levels. In the single-level AIM, this corresponds to spin degeneracy, which is lifted by applying a global magnetic field in the $z$-direction.

The Hamiltonian of a free electron gas (the environment) under the influence of a magnetic field in the $z$-direction now reads
\begin{equation}
    H_\text{env} \mapsto \sum_{k\sigma}\left[\epsilon_{\veck} + B S^z_{\veck} + \mu(B) \right]c^\dag_{\veck\sigma}c^{\phantom\dag}_{\veck\sigma} \ ,
\end{equation}
where we consider a single-level environment as in Eq.~\eqref{eq: hamiltonian anderson impurity multilevel multiorbit}, $B$ is expressed in units of energy, and $\mu(B)$ is the chemical potential, which adjusts such that the spin-up and spin-down species have the same chemical potential, in line with the equilibrium configuration. At $T=0$ and for a small magnetic field $B \ll \mathcal{D}$, assuming a constant density of states $d_0$ and a half-filled environment (as throughout this work), we obtain $\mu(B)=0$. Therefore, we can rewrite the single AIM, Eq.~\eqref{eq: hamiltonian anderson impurity multilevel multiorbit}, as
\begin{equation}
    H \to H + B\left( S^z_\text{imp} + S^z_\text{env}\right)  \ ,
\end{equation}
where the spin-up (spin-down) bath electrons have bandwith $\left[-\mathcal{D}+B,\mathcal{D}+B\right]$ $(\left[-\mathcal{D}-B,\mathcal{D}-B\right])$. The modified bandwidth has to be taken into account in the Wilson transformation, see App.~\ref{App: Wilsons transformation results} for a detailed discussion. With this, the Hamiltonian of the SIAM reads \begin{align}
    H = & H_\text{imp} + BS^z_\text{imp} + \sum_{\sigma,n=0}^\infty \epsilon_n^\sigma(B) c_{n\sigma}^\dag c_{n\sigma}^{\phantom\dag} \nonumber \\
     & + \sum_\sigma t_\text{D} \left( d_\sigma^\dag c_{0\sigma}^{\phantom\dag} + \text{H.c.} \right) \nonumber \\
      & + \sum_{\sigma,n=0} t_n^\sigma(B) \left(c_{n\sigma}^\dag c_{n+1\sigma}^{\phantom\dag}  + \text{H.c.}\right)
\end{align}
with spin-dependent Wilson chain coefficients $\epsilon_n^\sigma(B), t_n^\sigma(B)$.

In Fig.~\ref{fig: figure 3}(a), we show the $\mathcal{M}_n$ correlators along the Wilson chain for the SIAM in the presence of a finite magnetic field. As in the zero-field case, see Appendix~\ref{App: correlator along the Wilson chain}, the correlators in the Kondo regime initially increase with $n$ before decaying toward zero. This behavior is consistent with what is observed at zero magnetic field, as discussed in App.~\ref{App: correlator along the Wilson chain}. For sufficiently large magnetic fields, however, the large-$n$ coefficients no longer vanish; instead, they exhibit staggered oscillations around a finite background value. This reflects the coexistence of a uniform spin polarization induced by the field and the intrinsic alternating component of the bath spin correlations, as in the $B = 0$ case, see App.~\ref{App: correlator along the Wilson chain} for further details.

In Fig.~\ref{fig: figure 3}(b), we observe that $\mathcal{M}_\text{imp} \approx 1$ regardless of $B$. For $B < 10^{-5}U$, the environment almost perfectly screens the impurity, with $\mathcal{M}_\text{imp} \approx -\mathcal{M}_\text{env}$, and the impurity exists in a coherent superposition of the two spin configurations $\lambda_1 \approx \lambda_2 \approx 1/\sqrt{2}$. In this regime, the Kondo singlet still forms. For $B \lesssim 10^{-4}U$, we observe $\mathcal{M}_\text{env}>0.8$, indicating strong screening of the environment. The two nonzero Schmidt coefficients $\lambda_1, \lambda_2$ indicate that the system is not in a product state and that an impurity-environment hybridization persists. For $B\gtrsim 10^{-4}U$, the screening drastically decreases and the system is no longer hybridized $\lambda_1\to 1$, marking Zeeman splitting on the impurity. Crucially, our correlator captures the suppression of the Kondo effect. We note that the Kondo effect is suppressed at magnetic fields of similar order of magnitude than the Kondo gap, $B < k_\text{B} T_\text{K}$, as estimated in Eq.~\eqref{eq: Kondo temperature}.

\subsection{Two-level AIM}\label{Sec: Two-level AIM}
\begin{figure*}
\centering\includegraphics[width=17.4cm]{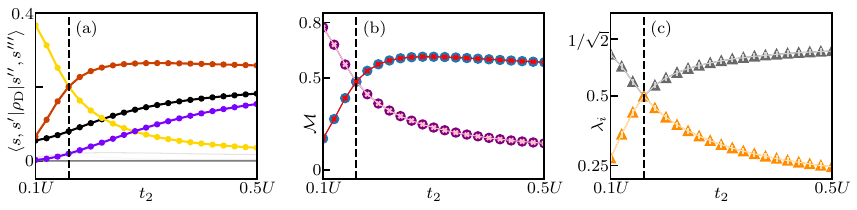}
	\caption{\textit{Kondo impurities and their screening in the two-channel AIM, cf. Eq.~\eqref{eq: hamiltonian anderson impurity multilevel multiorbit}.} (a) Tomography of the impurity's reduced density matrix, where we highlight the density matrix elements $\bra{0,0} \rho_\text{imp} \ket{0,0}$ (black), $\bra{0,\sigma} \rho_\text{imp} \ket{0,\sigma}$ (red), $\bra{\sigma,0} \rho_\text{imp} \ket{\sigma,0}$ (yellow), and $\bra{0,\uparrow\downarrow} \rho_\text{imp} \ket{0,\uparrow\downarrow}$ (purple). We depict  spin-degenerate configurations with $\sigma\in\left\{\uparrow,\downarrow\right\}$ via a single line, and plot the remaining elements  $\left|\bra{s,s'} \rho_\text{imp} \ket{s'',s'''}\right|< 0.1$ using thin grey lines. (b) $\mathcal{M}_\ell$ correlator of the impurity's level $\ell=1,2$ (purple and blue circles, respectively), cf.~Eq.~\eqref{eq: total projected spin impurity}, and $\mathcal{M}_{\text{env}_\ell}^\ell$ correlators of the environment's level $\ell=1,2$ (pink and red crosses, respectively). (c) Schmidt decomposition $\lambda_i$ with $i=1,\ldots,4$, (dark grey, light grey, dark orange, light orange, respectively) as in~Eq.~\eqref{eq: schmidt decomposition imp-env}. We set the impurity's onsite energy $\epsilon_\ell = -U/2$, impurity-environment coupling $t_1 = 0.2U$, assume the environment to have a constant density of states $d_0 = 1/(2U)$, set Wilson discretization parameter $\Lambda = 2$, Wilson chain length $N_\text{chain}^\ell = 100$, and MPS bond dimension $D = 300$. The dashed vertical lines mark the $t_1 = t_2$ point.}
	\label{fig: figure 4}
\end{figure*}
As a third example, we consider the two-level AIM, cf.~Eq.~\eqref{eq: hamiltonian anderson impurity multilevel multiorbit}, with $\ell \in \{1, 2\}$. We focus on the regime where $\epsilon_\ell = -U/2$ and the impurity Hamiltonian has a four-fold degenerate ground state
\begin{equation}
\ket{\sigma, 0} \quad \text{and} \quad \ket{0, \sigma} \ ,
\end{equation}
with $\sigma \in \{\uparrow, \downarrow\}$. The state $\ket{s, s'} = \ket{s}_1 \otimes \ket{s'}_2$ denotes the state of the first and second levels. Before applying our method, we use the tomography procedure to explore if, and in which form, the impurity hybridizes with the environment, see Fig.~\ref{fig: figure 4}(a). When the impurity-environment couplings of the two levels are equal, $t_1 = t_2$, the system reduced density matrix is approximately
\begin{equation}
\rho_\text{imp} = \frac{1}{4} \sum_\sigma \left( \ket{0, \sigma}\bra{0, \sigma} + \ket{\sigma, 0}\bra{\sigma, 0} \right) \ ,
\end{equation}
which corresponds to the impurity having been in a Kondo $SU(4)$ configuration. For $t_1 \neq t_2$, the $SU(4)$ configuration is suppressed, and the impurity’s reduced density matrix for $t_1 > t_2$ reads
\begin{equation}
\rho_\text{imp} = \frac{1}{2} \sum_\sigma \ket{\sigma, 0} \bra{\sigma, 0} \ ,
\end{equation}
while for $t_1 < t_2$ it is
\begin{equation}
\rho_\text{imp} = \frac{1}{2} \sum_\sigma \ket{0, \sigma} \bra{0, \sigma} \ .
\end{equation}
These reduced density matrix configurations correspond to a Kondo SU(2) in level 1 of the impurity and a Kondo SU(2) in level 2 of the impurity, respectively. This abrupt suppression of Kondo $SU(4)$ was theoretically identified using a NRG study~\cite{kleeorin_abrupt_2017,lopes_su4-su2_2017}. In the case of strong $t_2 \gg t_1$, the impurity's density matrix corresponds to a Kondo $SU(2)$ coexisting with the empty and double occupied configuration. As discussed above, while the tomography procedure reveals different types of impurity-environment hybridizations, it does not conclusively imply the formation of Kondo screening. Therefore, the tomography results act only as witnesses to the formation of Kondo screening.

To determine if the predicted Kondo $SU(2)$ and Kondo $SU(4)$ impurity states actually form, we apply our method to calculate the $\mathcal{M}_\ell, \mathcal{M}_{\text{env}_{\ell'}}^\ell$ correlators. In Fig.~\ref{fig: figure 4}(b), we observe near-perfect screening of the $\ell^\text{th}$ impurity's level magnetic moment, with $\mathcal{M}_{\ell} \approx -\mathcal{M}_{\text{env}_\ell}^{\ell}$, while $\mathcal{M}_{\text{env}_{\ell'}}^{\ell}\to 0$ for $\ell\neq\ell'$ (not plotted). In Fig.~\ref{fig: figure 4}(c), we plot Schmidt coefficients as in Eq.~\eqref{eq: schmidt decomposition imp-env}. For $t_1 > t_2$, we observe $\mathcal{M}_{1} \approx 0.8 \gg \mathcal{M}_{2} $ and nonvanishing $\lambda_i \approx 1/\sqrt{2}$ for $i=1,2$. This indicates that a Kondo $SU(2)$ effect forms with level 1. For $t_2 > t_1$, we observe $\mathcal{M}_2 \approx 0.6 \gg \mathcal{M}_{1}$ and nonvanishing Schmidt coefficients $\lambda_i \approx 1/\sqrt{2}$ for $i=1,2$. This indicates that a Kondo $SU(2)$ effect forms in level 2, coexisting with the empty and doubly occupied configuration. For $t_1=t_2$, we observe $\mathcal{M}_1\approx \mathcal{M}_2\approx 0.5$ and Schmidt coefficients $\lambda_i \approx 1/2$ for $i=1,2,3,4$, in line with the formation of a Kondo $SU(4)$ hybridization. Thus, we conclude that it is possible to directly identify the formation of Kondo $SU(N)$ effects by calculating the $\mathcal{M}_{\ell}, \mathcal{M}_{\text{env}_{\ell'}}^{\ell}$ correlators.

\section{Discussion}\label{sec: conclusion and outlook}
Our study, thus far, focused on a sub-class of impurity problems, i.e. multilevel impurities as in Eq.~\eqref{eq: hamiltonian anderson impurity multilevel multiorbit}. We postulate now a possible generalization of our method to multichannel, multiorbital Kondo impurities. The Hamiltonian of the latter reads
\begin{equation}\label{eq: Hamiltonian Kondo multichannel multilevel}
	{H} = \sum_{m\veck\alpha}\epsilon_{m\veck}^{\phantom\dag} c^\dag_{m\veck\alpha}c^{\phantom\dag}_{m\veck\alpha} + J \sum_{A=1}^{N^2-1} \sum_{mm'\veck \veck'} \boldsymbol{S}_\text{D}^A \boldsymbol{S}_{m \veck, m' \veck'}^A \ ,
\end{equation}
where $m=1,\ldots M$ represents the channel index and $\alpha = 1,\ldots,N$ denotes the spin index. As such, the impurity has an $SU(N)$ symmetry, whose fundamental representation has $A=1,\ldots,N^2-1$ generators. The state of the full system in this configuration is given by
\begin{equation}
\ket{\psi} = \frac{1}{\sqrt{\sum_\alpha \left|a_\alpha\right|^2}}\sum_\alpha a_\alpha\ket{\alpha}_\text{imp}\otimes\ket{\alpha'}_\text{env} \ ,
\end{equation}
where $\ket{\alpha}_\text{imp}$ is the impurity's state with spin $\alpha$ and the $\left\{\ket{\alpha'}\right\}$ are a set of orthogonal states, whose form is generally unknown. Note that this state differs from the one described in Eq.~\eqref{eq: Kondo state impurity plus environment multilevels}, as the impurity here consists of a single energy level with $N \geq 2$ orbital states. These orbital states can be screened by an environment comprising multiple flavors (channels). This multichannel Kondo model can exhibit complex screening mechanisms depending on the channel configurations and coupling strengths~\cite{sengupta_overscreened_1996,parcollet_transition_1997,parcollet_overscreened_1998}. In such a scenario, the environment states do not always perfectly screen the magnetic impurity. We define perfect screening if
\begin{equation}\label{eq: multichannel Kondo full screening condition}
    \bra{\alpha}S_\text{imp}^z\ket{\alpha}_\text{imp} =  -\bra{\alpha'}S_\text{env}^z\ket{\alpha'}_\text{env} \ .
\end{equation}

As such, we can generalize our $\mathcal{M}_{\ell}, \mathcal{M}_{\text{env}_{\ell'}}^{\ell}$ correlators to identify the presence of such Kondo screening mechanisms. To achieve this, the operators defined in Eq.~\eqref{eq: projector operators spin subhilbert spaces} can be adapted to project the system into the $\alpha$ subspaces of the impurity. The total spin of the impurity and the environment is then calculated following the approach outlined in Eqs.~\eqref{eq: total projected spin impurity} and~\eqref{eq: total projected spin environment}.

\section{Conclusion and outlook}
We have introduced a straightforward yet powerful methodology for capturing the role of the environment in screening Kondo impurities. This approach is particularly well-suited for studies involving impurities using the NRG-MPS technique, and more broadly for any impurity system formulated within the MPS framework. We have demonstrated its accuracy through results from the single AIM, where we observe perfect Kondo $ SU(2) $ screening and its suppression via a magnetic field, as well as the two-level AIM. In the latter, we detect Kondo screening in the crossover between $ SU(2) $ and $ SU(4) $ Kondo effects. These effects were reported on in the literature via other methods~\cite{kleeorin_how_2019}. Our results show exceptional agreement in the captured screening, which suggests that our method holds promise for studying screening mechanisms in more complex Kondo impurities, which are challenging to study via standard NRG~\cite{toth_density_2008,kugler_strongly_2020,stadler_hundness_2019}.

Our method is very versatile and can be applied to a variety of contemporary impurity studies. For example, a recently proposed algorithm called ``fork tensor network product state method''~\cite{bauernfeind_fork_2017} provides a powerful tool for extending the study of multichannel, multiorbital impurity systems. This algorithm efficiently encodes impurity systems coupled to multiple environments within the MPS formalism, making it directly applicable for calculating the $\mathcal{M}_{\ell}, \mathcal{M}_{\text{env}_{\ell'}}^{\ell}$  correlators proposed here. Additionally, our approach is not limited to ground-state calculations and can  be applied to open systems, where the system's configuration is described by an MPO instead of an MPS~\cite{jaschke_one-dimensional_2018,van2018entanglement}. This extension requires a finite-temperature, out-of-equilibrium MPS-based quantum impurity solver, with several recent proposals addressing this challenge~\cite{lotem_renormalized_2020, thoenniss_efficient_2023,kloss_equilibrium_2023,park_tensor_2024,cao_finite-temperature_2024}.

\begin{acknowledgements} We acknowledge financial support from the Swiss National Science Foundation (SNSF) through project 190078 and through Sinergia Grant No. CRSII5 206008/1, as well as from the Deutsche Forschungsgemeinschaft (DFG) - project number 449653034. We thank the constructive and educational Referees at PRB for their insightful and professional comments, which assisted us in improving this manuscript. Our numerical MPS implementations are based on the ITensors \textsc{Julia} library~\cite{fishman_itensor_2022}.
\end{acknowledgements} 

\section*{Data Availability}
The data that support the findings of this article are openly
available~\cite{data_repository}.
\appendix
\section{Recursive formula of the Wilson transformation}\label{App: Wilsons transformation results}
In this Appendix, we provide a concise description of Wilson's transformation, a key step in the NRG approach. We point the reader to Ref.~\cite{bulla_numerical_2008} for a detailed pedagogical review. Our discussion focuses on the essential parameters required for single AIM, as defined in Eq.~\eqref{eq: hamiltonian anderson impurity multilevel multiorbit}. We define the hybridization function
\begin{equation}
    \Gamma(\epsilon) = \pi \sum_{\veck} t_{\veck}^2 \delta(\epsilon_{\veck} - \epsilon) \ ,
    \end{equation}
   which is assumed to be zero outside the $\left[-\mathcal{D},\mathcal{D}\right]$ bandwidth. The first step of the Wilson's transformation involves discretizing this bandwidth by defining the discretization points $x_n = \pm \mathcal{D}\Lambda^{-n}$, where $\Lambda >1$ is the discretization parameter of the NRG transformation. In the following, we define
    \begin{equation}
	\int^{\pm,n} d\epsilon \equiv \pm\int_{\pm x_{n+1}}^{\pm x_n}d\epsilon \ .
\end{equation} 
    With this, we define multiple variables which will be inserted in the recursive relation including
    \begin{subequations}\label{Eq: NRG Bulla coefficients}
    \begin{align}
    \eta_0 & = \int_{-\mathcal{D}}^\mathcal{D}d\epsilon\Gamma(\epsilon)\epsilon \ , \\
	\xi_n & = \frac{\int^{\pm,n} d\epsilon\Gamma(\epsilon)\epsilon}{ \int^{\pm,n} d\epsilon\Gamma(\epsilon)} \ , \\
\left(\gamma_n^{\pm}\right)^2 & =\int^{\pm,n}d\epsilon\Gamma(\epsilon)   \ . 
\end{align}
\end{subequations}
After some intermediate steps, the single AIM, cf. Eq.~\eqref{eq: hamiltonian anderson impurity multilevel multiorbit} for $\ell\equiv 1$,
\begin{align}
    H  = & H_\text{imp} + \sum_{n\sigma}\left(\xi_{n}^+a^\dag_{n\sigma}a_{n\sigma}^{\phantom\dag} +\xi_n^-b^\dag_{n\sigma}b_{n\sigma}^{\phantom\dag}\right) \nonumber \\
    & + \frac{1}{\sqrt{\pi}}\sum_\sigma d_\sigma^\dag\sum_n\left(\gamma_n^+ a_{n\sigma}^{\phantom +} + \gamma_n^-b_{n\sigma}^{\phantom -} \right)  + \text{H.c.} \ ,
\end{align}
Where the $a_{n\sigma} (b_{n\sigma})$ are annihilation operators associated with the $n^\text{th}$ positive (negative) energy interval $[x_{n+1},x_n]$ ($[-x_n,-x_{n+1}]$). A semi-unitary transformation 
\begin{subequations}\label{eq: unitary transformation Wilson chain}
\begin{align}
a_{n\sigma} & = \sum_m^\infty u_{mn}c_{m\sigma} \\
b_{n\sigma} & = \sum_m^\infty v_{mn}c_{m\sigma} \\
     c_{n\sigma} & = \sum_m ^\infty u_{nm} a_{m\sigma} + v_{nm}b_{n\sigma} 
\end{align} 
\end{subequations}
maps this Hamiltonian to the form as in Eq.~\eqref{eq: hamiltonian AIM wilson chain}.
The energy and coupling coefficients are calculated with a recursion relation with initial values
\begin{subequations}
	\begin{align}
		& \epsilon_0  = \frac{1}{\eta_0}\int_{-\mathcal{D}}^{\mathcal{D}}d\epsilon \Gamma(\epsilon)\epsilon \ , \label{eq: wilson transformation e0} \\
		& t_0^2 = \frac{1}{\eta_0}\sum_m\left[\left(\xi_m^+ -\epsilon_0\right)^2\left(\gamma_m^+\right)^2 + \left(\xi_m^- -\epsilon_0\right)^2\left(\gamma_m^-\right)^2\right] \ , \label{eq: wilson transformation t0} \\
        & u_{0m}  = \frac{\gamma^+}{\sqrt{\eta_0}} \ , \\
     & v_{0m}  =  \frac{\gamma^-}{\sqrt{\eta_0}} \ , \\
		& u_{1m}  = \frac{1}{t_0}\left(\xi_m^+ -\epsilon_0 \right)u_{0m} \ , \label{eq: wilson transformation u1m}  \\
		& v_{1m} = \frac{1}{t_0}\left(\xi_m^- -\epsilon_0 \right)v_{0m} \ . \label{eq: wilson transformation v1m}  
	\end{align}\label{eq: wilson transformation starting values} 
\end{subequations}
Subsequent values are calculated as
\begin{subequations}
	\begin{align}
		&\epsilon_n =  \sum_m \left(\xi_m^+u_{nm}^2 + \xi_m^-v_{nm}^2 \right) \ , \label{eq: wilson transformation en} \\
		&t_n^2 =  \sum_m \left[\left(\xi_m^+\right)^2u_{nm}^2 + \left(\xi_m^-\right)^2v_{nm}^2\right] -t_{n-1}^2 -\epsilon_n^2\ , \label{eq: wilson transformation tn} \\
		&u_{(n+1)m}  = \frac{1}{t_n}\left[\left(\xi_m^+ -\epsilon_n \right)u_{nm} - t_{n-1}u_{(n-1)m}\right] \ , \label{eq: wilson transformation unm}  \\
		&v_{(n+1)m}  = \frac{1}{t_n}\left[\left(\xi_m^+ -\epsilon_n \right)v_{nm} - t_{n-1}v_{(n-1)m}\right] \ , \label{eq: wilson transformation vnm}  
	\end{align}\label{eq: wilson recursion}\end{subequations}where $\epsilon_n$ is the on-site energy amplitude of the $n^\text{th}$ site of the Wilson chain and $t_n$ is the hopping amplitude between the $n^\text{th}$ and $(n+1)^\text{th}$ chain site, cf. Eq.~\eqref{eq: hamiltonian AIM wilson chain}.
    In this work, we consider an energy-dependent hybridization function $\Gamma(\epsilon) = const.$. With this assumption, at zero magnetic field $B=0$ the energy and hopping coefficients of thw Wilson chain result
    \begin{equation}\label{eq: simplification of Wilson transf for constant hybridization function}
        \epsilon_n = 0, \ \ \ u_{nm} = (-1)^nv_{nm} \ .
    \end{equation}
    For a finite magnetic field $B\neq 0$, the bath energy band becomes $\left[-\mathcal{D}+B,\mathcal{D}+B\right]$ ($\left[-\mathcal{D}-B,\mathcal{D}-B\right]$) for spin-up (spin-down) electrons. Consequently, the discretization intervals $x_n$ and the integral limits in Eqs.~\eqref{Eq: NRG Bulla coefficients} must be adapted to this asymmetric band structure, splitting negative and positive energies around zero. With these modifications, one obtains $\epsilon_0^{\sigma} = \sgn(\sigma) B$ as in Eq.~\eqref{eq: wilson transformation e0}. The remaining coefficients follow from the usual recursive construction applied separately to each spin species, yielding a Hamiltonian with $\sigma$-dependent parameters $\epsilon_n^\sigma$ and $t_n^\sigma$.

\section{$\mathcal{M}_n$ correlator along the Wilson chain}\label{App: correlator along the Wilson chain}
\begin{figure}
\centering\includegraphics[width=8.6cm]{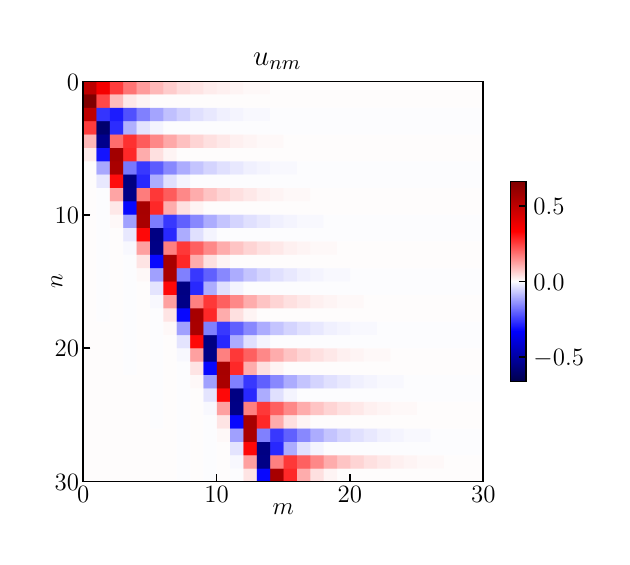}
	\caption{\textit{Graphical representation of the Wilson transformation.} Plot of the coefficients of the semi-unitary Wilson transformation, cf. Eq.~\eqref{eq: unitary transformation Wilson chain} for $\Lambda=2\mathcal{D}$ and constant hybridization function $\Gamma=const.$. For this choice, $v_{nm} = (-1)^n u_{nm}$. We set $N_\text{chain} = 30$ for illustration; the specific value of the chain length is arbitrary.}
	\label{fig: figure S1}
\end{figure}
In Figs.~\ref{fig: figure 2}(b) and~\ref{fig: figure 3}(a), we plot the correlator $\mathcal{M}_n$ for each site of the Wilson chain. We observe an oscillatory behavior between even and odd chain sites and, in Kondo regimes, a growth of the absolute value of the local correlator before it decays to a constant (zero at $B = 0$). A complete analysis of the screening mechanism would in principle require a ``backward Wilson transformation'' involving the evaluation of all correlators. This, however, would obscure the main advantage of the Wilson transformation: reducing the star geometry of the impurity problem to a chain geometry. Our method takes advantage of this simplification, since it requires only the expectation value of diagonal operators to extract order parameters (the $\mathcal{M}_\text{imp}$ and $\mathcal{M}_\text{env}$ correlators). Nevertheless, the physics of screening is qualitatively understood and the Wilson transformation itself has well-defined mathematical properties. In the following, we connect these two aspects to develop an intuition for the observed behavior.

First, we note that the alternating structure of $M_n$ has a direct physical interpretation. The impurity perturbs the bath locally, and the induced spin response at site $n$ is proportional to the bath spin-spin correlation function. In a fermionic 1D bath, the local spin density operator contains an oscillatory component at the Fermi wavevector $2k_F$, arising from interference between left- and right-moving states near the Fermi points~\cite{affleck_critical_1991,affleck_friedel_2008}. 
Consequently, the bath correlation function contains a staggered contribution proportional to $\cos(2k_F n)$, which dominates the long-distance behavior. In this work, we consider the half-filling regime $2k_F=\pi$, resulting in the alternating factor $(-1)^n$. Note that the Wilson index $n$ parametrizes logarithmic energy scales rather than real-space positions, the oscillations of $M_n$ reflect how screening weight is distributed across successive energy shells. Additionally, at finite magnetic field $B\neq 0$, the correlator develops a disconnected contribution $\langle S^z\rangle^2$, producing oscillations around a finite constant background.

Second, we note that the initial growth of the amplitude signals the onset of hybridization between the impurity and the conduction electrons, leading to the formation of the screening cloud, while the subsequent decay at larger $n$ reflects its finite spatial extent (lower energies correspond to longer length scales).

Lastly, we consider the asymmetric even-odd behavior. This behaviour requires a more careful analysis and can be traced back to the structure of the Wilson transformation. To make this connection explicit, we now turn to the calculation of $\mathcal{M}_n$, which involves a projection followed by the evaluation of the $S^z_{n}$ operators. In terms of the operators of the Wilson chain, the latter reads
\begin{equation}
    S_n^z = \sum_\sigma \bra{\sigma}S^z\ket{\sigma} c^{\dag}_{n\sigma}c^{\phantom\dag}_{n\sigma} \ .
\end{equation}
In the discretized basis and assuming a constant hybridization function, cf. Eqs.~\eqref{eq: wilson recursion} and~\eqref{eq: simplification of Wilson transf for constant hybridization function}, the occupation operators take the form
\begin{align}\label{eq: occupation operators in the discretized basis}
   & c^{\dag}_{n\sigma}c^{\phantom\dag}_{n\sigma} 
    = \sum_{m,m'=0}^\infty u_{nm}u_{nm'} \Biggl[
        a_{m\sigma}^\dag a_{m'\sigma}^{\phantom\dag} \nonumber \\ 
       & + (-1)^n\!\left(a_{m\sigma}^\dag b_{m'\sigma}^{\phantom\dag}
        + b_{m\sigma}^\dag a_{m'\sigma}^{\phantom\dag}\right)
        + b_{m\sigma}^\dag b_{m'\sigma}^{\phantom\dag}
    \Biggr] \ .
\end{align}
Since the $u_{nm}$ coefficients define a unitary transformation, the sum over all local $S_n^z$ operators results in
\begin{equation}
    \sum_n S_n^z = \sum_{n\sigma} \bra{\sigma}S^z\ket{\sigma} \left(a^{\dag}_{n\sigma}a^{\phantom\dag}_{n\sigma} + b^{\dag}_{n\sigma}b^{\phantom\dag}_{n\sigma}\right) \ .
\end{equation}
Hence, to compute the expectation value in the original (discretized) star-geometry Hamiltonian, it suffices to sum the corresponding values in the transformed Wilson chain. 

From Eq.~\eqref{eq: occupation operators in the discretized basis}, we see that the $\mathcal{M}_n$ values depend on the coefficients $u_{nm}$ as well as on the expectation values of two-point correlators projected onto different (positive for particles and negative for antiparticles) energy intervals
\begin{equation}
    \bra{\psi}P^\sigma a^\dag_{m\sigma} a^{\phantom\dag}_{m'\sigma} P^\sigma\ket{\psi}  \ , \ \   \bra{\psi}P^\sigma a^\dag_{m\sigma} b^{\phantom\dag}_{m'\sigma} P^\sigma\ket{\psi} \ ,
\end{equation}
and similarly for $a \mapsto b$, $b\mapsto a$. The values of the two-point correlators in the projected space are directly linked to the energy intervals involved in the transformation and to how the spins in these intervals screen the Kondo impurity. 
 
The $u_{nm}$ coefficients have distinct mathematical properties, as can be observed in Fig.~\ref{fig: figure S1}. Specifically, they reach a maximum value with a tail and appear pairwise with positive $n=0,1,4,5,8,9,\ldots$ and negative $n=2,3,6,7,10,11,\ldots$ coefficients. From Eq.~\eqref{eq: occupation operators in the discretized basis}, we observe that in the calculation of the $S_z^n$ operators the prefactor $u_{nm}u_{nm'}$ is positive for any $n$ and sufficiently large $m,m'$. Additionally, we note a distinction between even and odd Wilson chain sites, where the tail of the odd $n$ is considerably shorter than that of the even sites $n$. This property, together with the $(-1)^n$ prefactors of the terms mixing particles and antiparticles, explains the asymmetry in the alternating even/odd coefficients.

\end{document}